\documentclass[mathleft]{an}
\usepackage{graphicx}
\usepackage{times}
\begin{document}
\def\teff{$T_{\rm eff}$}
\def\kms{{km~s}$^{-1}$}
\def\logg{$\log g$}
\def\turb{$\xi$}
\def\rad{$V_{\rm r}$}
\def\elem{$\varepsilon ({\rm El})$}
\def\vsini{$V\sin i$}

\Pagespan{789}{}
\Yearpublication{2010}%
\Yearsubmission{2010}%
\Month{11}%
\Volume{999}%
\Issue{88}%

\title{Pulsational content and abundance analysis of some $\delta$ Scuti stars \\ observed by CoRoT 
\thanks{
The CoRoT space mission was developed and  is operated by the French
space agency CNES, with participation of ESA's RSSD and Science Programmes,
Austria, Belgium, Brazil, Germany, and Spain.
This work is based on observations made with ESO Telescopes at the La Silla Observatory under  
the ESO Large Programmes LP 178.D-0361, LP 182.D-0356, and LP 185.D-0056.}}
\author{E. Poretti\inst{1}\fnmsep\thanks{\email{ennio.poretti@brera.inaf.it}},
L. Mantegazza\inst{1},
E. Niemczura\inst{2},
M. Rainer\inst{1}, 
T. Semaan\inst{3},
T. L\"uftinger\inst{4},
A. Baglin\inst{5}, 
M. Auvergne\inst{5}, E. Michel\inst{5}  \and R. Samadi\inst{5}
}
\titlerunning{CoRoT and Delta Sct stars}
\authorrunning{E. Poretti  et al.}
\institute{
INAF-Osservatorio Astronomico di Brera, Via E. Bianchi 46, 23807 Merate, Italy
\and 
Astronomical Institute of the Wroclaw University, ul. Kopenika 11, 51-622 Wroclaw, Poland
\and
GEPI, Observatoire de Paris, CNRS, Universit\'e Paris Diderot,  92195 Meudon, France
\and
Institut f\"ur Astronomie, Universit\"at Wien, Turkenschanzstr. 17, A-1180 Wien, Austria
\and
LESIA, UMR8109, Universit\'e Paris Diderot, Observatoire de Paris, 92195 Meudon, France
}

\received{01 April 2010}
\accepted{--}
\publonline{later}

\keywords{stars: abundances, stars: oscillations, stars: individual (HD50844, HD 50870, HD 292790, 
HD 171586, HD 172588, HD 1712748), delta Scuti stars, stars: rotation}

\abstract{Several $\delta$ Sct stars were photometrically monitored with the satellite CoRoT and observed in high--resolution
spectroscopy   from ground. We present here the preliminary analysis of the abundance analysis of several potential CoRoT
targets. Moreover, new insights are given about the problem of the huge number of detected frequencies by comparing
$\delta$ Sct stars with normal A-stars.}

\maketitle

\section{Combining space photometry and ground--based spectroscopy}

The ground--based campaigns on  $\delta$ Sct stars revealed richer and richer frequency spectra of
these opacity--driven pulsators.
The  continuous improvements in their exploitation
suggested that the detection of a wealth of very low amplitude
modes was only a matter of signal--to--noise ratio (Garrido \& Poretti 2004).
The confirmation of this suggestion was offered by
the space mission CoRoT (COnvection, ROtation and planetary Transits; Baglin et
al. 2006).
As a matter of fact, the frequency analysis of the CoRoT timeseries on HD~50844 
(Poretti et al. 2009), the first $\delta$ Sct star observed by CoRoT in the
Initial Run, revealed hundreds of terms in the  frequency range 0--30~d$^{-1}$. 

\begin{figure}
\centering
\includegraphics[width=0.45\textwidth]{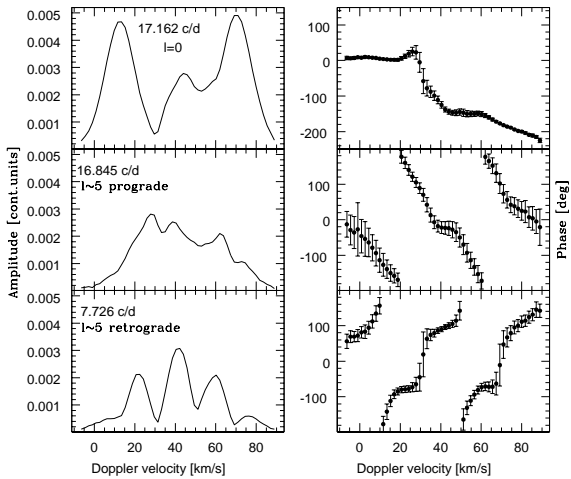}
\caption{Amplitude and phase plots of three modes observed in the spectroscopic
data of the $\delta$ Sct star HD~50870.}
\label{lm}
\end{figure}

The mode identification of  solar-like oscillators rely on the
regularity of the frequency pattern in the power spectrum. This method
may not be applicable to larger amplitude
pulsators, as $\delta$ Sct stars, for which non--linear effects select the observed modes in a
way which is not yet fully understood. For these targets, the
identification of modes observed by CoRoT may present some difficulties,
and will require additional data collected  from the ground.
In particular, ground-based spectroscopic observations are needed to assign
$m$-values to the detected modes and 
low--degree modes  have to be complemented with the
observation of the  high--degree ones. Moreover, 
diagnostics of stellar  interiors are often strongly
weighted toward the surface; one example is the rotational splitting, for which the
structure and rotation of sub--surface regions play a dominant role.
A precise and reliable
knowledge of sub--surface layers can be extracted from high--degree modes
and these modes can be typified by high-resolution spectroscopy only.
Figure~\ref{lm} shows the case of the $\delta$ Sct star HD~50870
(Mantegazza et al., in preparation). The amplitude
and phase plots suggest the identification 
of the displayed modes as radial (top panels), prograde non-radial (middle
panels), and retrograde non-radial (bottom panels). 

In the specific case of HD 50844, the spectroscopic data
allowed us to identify  high--degree modes up to $\ell=14$.
As an immediate consequence, we were forced to admit that the  cancellation effects
are not sufficient in removing the flux variations associated to these modes at the
noise level of the CoRoT measurements.
Moreover, the predominant term ($f_1$=6.92~d$^{-1}$) was typified as the  
fundamental radial mode.

\begin{figure}
\centering
\includegraphics[width=0.45\textwidth]{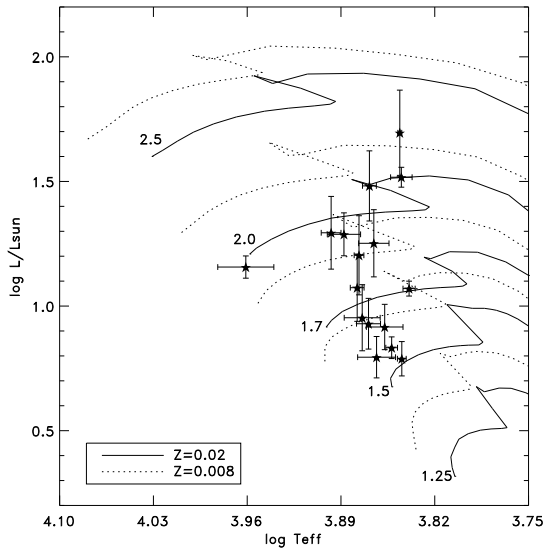}
\caption{
Hertzsprung-Russel diagram for sample of $\delta$\,Scuti stars from the
CoRoT field of view. The evolutionary tracks for $Z=0.02$ (solid lines)
and $Z=0.008$ (dashed lines) were taken from Lejeune \& Schaerer (2001).
}
\label{CMD}
\end{figure}

\begin{figure}
\centering
\includegraphics[width=0.45\textwidth]{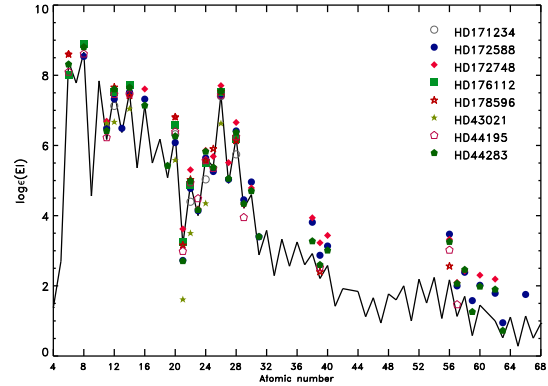}
\caption{
The elemental abundances of potential CoRoT targets of $\delta$\,Scuti
type in comparison with the solar abundances (solid line) from Grevesse
et al. (2007). 
 }
\label{f3}
\end{figure}

\section{Abundance analysis of CoRoT $\delta$ Sct  stars}
The ground-based observations indicate that HD\,50844 is an evolved star that is slightly 
underabundant in heavy elements, located on the Terminal Age Main Sequence. The CoRoT 
scientific programme contains other $\delta$\,Sct stars, with different evolutionary 
statuses, and a careful abundance analysis can greatly help in the selection process.

Our analysis follows the methodology presented in Niemczura \& Po\l{}ubek (2006) and 
relies on an efficient spectral synthesis based on a least-squares optimisation algorithm 
(see Takeda 1995; Bevington 1969). The method allows for the simultaneous determination of 
various parameters involved in stellar spectra and consists of the minimisation of the 
deviation between the theoretical flux distribution and the observed normalised one. 
The synthetic spectrum depends on the stellar parameters. 
The effective temperature $T_{\rm eff}$, surface gravity $\log g$, and
microturbulence $\xi$ were obtained before the determination of
abundances of chemical elements and were considered as input parameters.
The rotational velocity $v\sin i$, radial velocity $V_{\rm r}$, 
and the relative abundances of the elements $\epsilon \rm {(El)}$ can be determined 
simultaneously because they produce detectable and different spectral signatures. 
The $v\sin i$ values were determined by comparing the shapes of metal line profiles 
with the computed profiles, as shown in Gray (2005). The theoretical spectrum was 
fitted to the normalised observed one and the continuum was objectively drawn by 
connecting the highest points of the analysed spectrum part. For the adopted input 
values, the solution is typically reached after 10 iterations. The program stops 
if the values of the determined parameters remain the same within 2\% for three 
consecutive steps. The atmospheric models used in this analysis were computed with 
the line-blanketed LTE ATLAS9 code (Kurucz 1993). The synthetic spectra were 
obtained with the SYNTHE code (Kurucz 1993).

Figure\,2 illustrates the positions in the HR diagram for the candidate targets having 
HARPS and/or FEROS spectra. The effective temperature was determined by using both 
photometric and spectroscopic approaches. We used colours of three different photometries: 
Str\"{o}mgren {\it uvby$\beta$} (Moon \& Dworetsky 1985; Napiwotzki, Schoenberner \& Wenske 1993), 
Geneva (K\"{u}nzli et al. 1997) and $2MASS$ index $K$ (Masana, Jordi \& Ribas 2006). The 
spectroscopic methods included the determination of $T_{\rm eff}$ from shapes of 
Balmer lines and from lines of neutral iron, Fe\,I. The effective temperature was 
adjusted till there was no trend in the Fe\,I abundance versus excitation potential 
of the atomic level causing the line. The last method was used for stars with 
small projected rotational velocity ($v\sin i\leq$ 35\,km\,s$^{-1}$). On Fig.\,2 
the average values of effective temperature obtained from all used methods are shown. 
The luminosities were determined from the standard relation 
$\log {L/L_{\odot}} = 0.4*(M_V^{\odot}-M_V+BC_\odot+BC)$, where $M_V=V+5*\log(\pi)+5$. 
The $V$ magnitudes and  {\it Hipparcos} parallaxes (van Leeuwen 2007) were taken from 
SIMBAD database. Bolometric corrections ($BC$) are based on data summarized by Flower (1996). 
The solar values $BC_{\odot}=-0.07$ and $M_V^{\odot} = 4.81$ are adopted from 
Bessell, Castelli \& Plez (1998). The evolutionary tracks were taken from Lejeune \& Schaerer (2001) 
for two metallicities ($Z=0.020$ and $Z=0.008$) and for stellar masses 
1.25, 1.5, 1.7, 2.0 and 2.5$M_{\odot}$. The two stars located after 
TAMS are HD\,172588 and HD\,172748 ($\delta$\,Sct). Both these objects are giants 
with spectral types F0\,II-III and F2\,IIIp, respectively. The location of HD\,172588, 
the most evolved star on HR diagram, can be determined by the error in luminosity, 
resulting from the uncertainty of {\it Hipparcos} parallax. 
The presented errors of effective temperatures are the standard
deviations derived from all $T_{\rm eff}$ values obtained for a given
star.

On Figure \,3 the chemical abundances collected for the analysed objects are 
compared with the solar values taken from Grevesse, Asplund \& Sauval (2007). Almost for 
all stars the rare-earth and heavy elements are overabundant. This result can 
be caused by the small number of lines of these elements in the stellar spectra. 
Additionally, all these features are weak or are blends of lines of other elements. 
In most cases, the light and iron peak elements are in the error bars similar 
to the solar values. The peculiar abundances of HD\,172748 were confirmed. 
The detailed description of used methods and obtained results will 
be presented in Niemczura et al. (2010, in preparation).

\section{New insights into the pulsational content}
\begin{figure}
\centering
\includegraphics[width=0.45\textwidth]{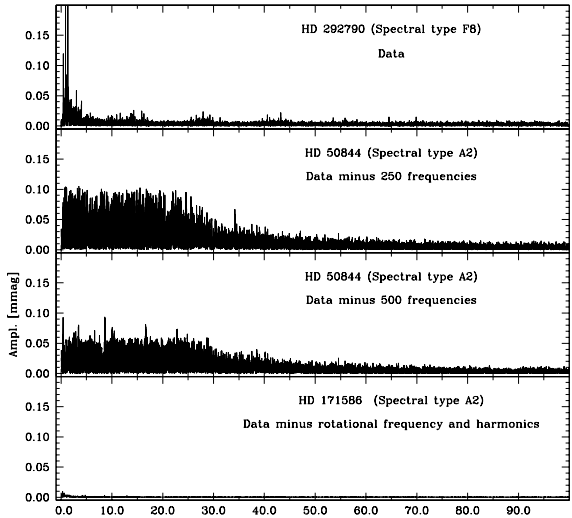}
\caption{Amplitude spectra of HD~292790 (top panel), HD 50844 (second and third panels), and
HD 171586 (bottom panel).}
\label{ampli}
\end{figure}
The potential contribution of granulation to the huge number of significant pics 
has been considered (Michel 2010 and also Kallinger \& Matthews 2010). 
Most of the peaks detected in the frequency spectrum of HD~50844 could be the 
signature of non--white granulation noise.
This possibility deserves attention and we bring here a new contribution.
Figure~\ref{ampli} shows the amplitude spectra of three stars.
In the top panel we see the spectrum of the rotational variable HD~292790. The
peaks at $f<$5~d$^{-1}$ are due to the rotational frequency and harmonics. There is
no other relevant peak, except those of the CoRoT orbital frequency (13.97~~d$^{-1}$)
and harmonics. The level of the noise is below 0.01~mmag. The two panels in the middle
show the amplitude spectra of the $\delta$ Sct star HD~50844 after subtracting 250 and
500 frequencies. The level of the peaks is 0.10~mmag in the second panel,
0.05~mmag in the third. The bottom panel shows the amplitude spectrum of another rotational
variable (HD~171856, L\"uftinger et al., in preparation). In this latter case the
rotational frequency and harmonics were subtracted and as a matter of fact  the spectrum shows the expected
white-noise at the level of 0.002~mmag. It is of interest to note that HD~171856
has the same spectral type of HD~50844 and it does not show any trace of noise due
to granulation. Therefore, it seems that the problem of the huge number of frequencies
detected in the amplitude spectra of $\delta$ Sct stars  is not solved yet,  since
noise excess is not observed in non--pulsating stars. Of course,
rotation and chemical composition near surface  can be  different and might influence
granulation.  However, if we admit that
granulation is contributing significantly to the excess of power below 30 ~d$^{-1}$, Fig.~\ref{ampli} suggests
that the onset of granulation effects is related to pulsation. It is quite evident that
the physical scenario of A--stars needs a careful revisitation to match the results obtained
from CoRoT observations.

\acknowledgements
We acknowledge the support from the {\it Centre National
d'Etudes Spatiales} (CNES).
The spectroscopic  data are being obtained as part of the ESO Large Programmes
LP178.D-0361, LP182.D-0356, and LP185.D-0056 (PI.: E.~Poretti) in the framework of  
the Italian ESS project, contract ASI/INAF I/015/07/0, WP 03170.
EP acknowledges the support from the European Helio and Asteroseismology
Network (HELAS) for the participation to the Conference.
EN acknowledges financial support of the N\,N203\,302635 grant from the MNiSW.


\begin{thebibliography}{}
\bibitem[Baglin et al.(2006)]{esa3} Baglin, A., Auvergne, M., Barge, P., et al., 2006,
in ``The CoRoT Mission, Pre-Launch Status,
Stellar Seismology and Planet Finding", M.~Fridlund, A.~Baglin, J.~Lochard, \& L.~Conroy, Eds.
(ESA SP-1306, ESA Publications Division, Nordwijk, Netherlands), p.~33
\bibitem[Bessell et al.(1998)]{bess} Bessell, M. S., Castelli, F., Plez, B., 1998, A\&A, 333, 231
 (Erratum: 1998, A\&A, 337, 321)
\bibitem[1969]{bevington} Bevington, P.R., 1969, in {\it Data reduction and error analysis for the physical sciences}, New York: McGraw-Hill
\bibitem[1996]{flower} Flower, P.J., 1996, ApJ, 469, 355
\bibitem[Garrido \& Poretti(2004)]{nz} Garrido, R., \& Poretti, E., 2004,
in ``Variable Stars in the Local Group", IAU Colloq.~193, D.W.~Kurtz \& K.R.~Pollard Eds., ASP Conf. Series, 310, 560
\bibitem[2005]{gray} Gray, D.F., 2005, {\it The Observation an Analysis of Stellar Photospheres}, 
Cambridge University Press, {\it The Edinburgh Building, Cambridge, CB2 2Ru, UK}
\bibitem[2007]{grevesse} Grevesse, N., Asplund, M., Sauval, A.J., 2007, SSRv, 130, 105
\bibitem[]{gran} Kallinger, Th., Matthews, J.M., 2010, ApJ, 711, L35
\bibitem[1997]{kun}K\"{u}nzli, M., North, P., Kurucz, R.L., Nicolet, B., 1997, A\&AS, 122, 51
\bibitem[1993]{cdrom} Kurucz, R., 1993, CD-ROM 18
\bibitem[2001]{lejeune} Lejeune, T., Schaerer, D., 2001, A\&A, 366,538
\bibitem[2010]{luft} L\"uftinger, T., et al.,  in preparation
\bibitem[2010]{man} Mantegazza, L., et al.,  in preparation
\bibitem[2006]{masana} Masana, E., Jordi, C., Ribas, I., 2006, A\&A, 450, 735
\bibitem[2010]{eric} Michel, E., 2010, these proceedings
\bibitem[1985]{moon}Moon, T., Dworetsky, M.M., 1985, MNRAS, 217, 305
\bibitem[1993]{napi} Napiwotzki, R., Schoenberner, D., Wenske, V., 1993, A\&A, 268, 653
\bibitem[2006]{niemczura} Niemczura, E., Po{\l}ubek, G., 2006, ESASP, 624, 120
\bibitem[2010]{niemc} Niemczura, E., et al.,  in preparation
 \bibitem[Poretti et al. (2009)]{Poretti09} Poretti, E., E. Michel, R. Garrido, et al., 2009, A\&A, 506, 85
\bibitem[1995]{takeda} Takeda, Y., 1995, PASJ, 47, 287
\bibitem[2007]{vl} van Leeuwen, F., 2007, A\&A, 474, 653
\end{thebibliography}
\end{document}